\documentclass[preprint2]{aastex}
\shorttitle{}
\shortauthors{}

\begin{document}

\title{An emission mechanism explaining off-pulse emission originating in the outer magnetosphere of pulsars.}

\author{Rahul Basu$^1$, Dipanjan Mitra$^1$ and George I Melikidze$^2$}
\affil{$^1$National Centre for Radio Astrophysics, P. O. Bag 3, Pune University
Campus, Pune: 411 007. India}
\affil{$^2$Kepler Institute of Astronomy, University of Zielona G\'{o}ra, 
Lubuska 2, 65-265, Zielona G\'{o}ra, Poland
}
\email{rbasu@ncra.tifr.res.in, dmitra@ncra.tifr.res.in, gogi@astro.ia.uz.zgora.pl}

\begin{abstract}
We have examined the cyclotron resonance instability developing in the 
relativistic outflowing plasma in the pulsar magnetosphere. The instability
condition leads to radio emission in the sub-GHz frequency regime which is 
likely to be seen as off-pulse emission. Recent studies have shown the 
presence of off-pulse emission in long period pulsars, and we demonstrate 
this plasma process to be an energetically viable mechanism. 
\end{abstract}


\section{\bf Introduction}
The cyclotron resonance instability in relativistic plasma which leads to
cyclotron maser emission has been proposed as a feasible mechanism for 
generating coherent emission at low radio frequencies in many astrophysical 
systems. The emission mechanism works in plasma systems where excess free 
energy in the particle distribution is unable to dissipate internally. The 
excess energy is directly emitted as electromagnetic radiation through a 
mechanism of negative absorption in the plasma, equivalent to a maser-like 
emission. The cyclotron resonance instability is expected to be the source 
of radio emission in a large number of space plasmas and astrophysical systems 
like the auroral kilometric radiation of the Earth, the auroral emission in 
Jupiter, Saturn and exoplanets, spikes in radio emission from Sun and stars 
at low radio frequencies, coherent radiation from blazar jets, and so on
\citep{tre06}.

The cyclotron resonance instability has been suggested as a possible mechanism
for the coherent main pulse emission in pulsars~\citep{kaz87,kaz91}.
Electromagnetic waves in the radio frequency regime, excited as a result 
of this emission mechanism, originate in the outer magnetosphere nearer to the 
light cylinder. However observations in the recent past have constrained the 
main pulse emission to originate near the neutron star surface at a height 
of $\sim$ 500 km, about 1\% the light cylinder distance
{\citep{ran93a,ran93b,kij98,kij03,mit02}}. This rules out the cyclotron 
resonance instability as a suitable emission mechanism for the main pulse.

Recent discoveries have shown the presence of coherent radio emission in the 
pulsar profile far away from the main pulse~\citep{bas11,bas12}. The coherent 
emission from pulsars is generated in the relativistic plasma flowing outward
from the stellar surface along the open magnetic field lines. The main pulse 
is observationally constrained to originate near the pulsar surface, thereby 
compelling any off-pulse emission far away from it to originate in the outer 
magnetosphere. In this paper we assess the plasma conditions necessary for 
the cyclotron resonance instability to develop in the outer magnetosphere 
and whether it can account for the detected off-pulse emission. In \S 2 we 
describe the general conditions in the pulsar magnetosphere leading to radio 
emission. \S 3 describes the condition for the cyclotron resonance instability 
to develop in the pulsar magnetosphere and its applicabilty to off-pulse 
emission. \S 4 deals with the energetics of the emission mechanism followed 
by a short discussion in \S 5 about the implications of this mechanism to 
pulsar radio emission.

\section{\bf Radio emission in Pulsars}
In this section we briefly describe the conditions in the pulsar magnetosphere
leading to radio emission. Most of the models explaining the emission mechanism 
involve the presence of an inner acceleration region above the polar cap 
leading to the formation of dense outflowing e$^-$e$^+$ plasma. We explore 
below the mechanism first described by~\citet{rud75}; hereafter RS75; and 
later modified by \citet{gil00} called the ``sparking" vacuum-gap model.

The neutron star, radius $R_{S}$ = 10$^6$ cm, is characterized by a strong 
magnetic field $B_s$, which is highly multipolar near the stellar surface. 
The dipole part of the surface magnetic field is given as $B_d$ = 3.2$\times
$10$^{19}~(P\dot{P}) ^{1/2}$ G, here $P$ and $\dot{P}$ are the rotation 
period and period derivative, respectively, $B_s$ = $bB_d$ and $b \gg$ 1. 
The rotating magnetic field gives rise to an electric field ($\bf{E}$), and 
in order to maintain the force-free condition ({$\bf{E\centerdot B}$} = 0) a 
corotating charge-separated magnetosphere, with density $n_{GJ} = 
-(\bf{\Omega\centerdot B})/2\pi ce$ \citep{gol69}, surrounds the neutron star. 
The force-free condition breaks down at the light cylinder, $R_{LC} = c/\Omega
\approx 4.8 \times 10^{9} P$~cm, beyond which the charges no longer corotate. 
Above the polar regions there is an outflow of charges and the magnetosphere 
slowly recedes, resulting in a charge depleted vacuum gap with a large 
potential drop across it. In the gap region $e^-e^+$ pairs are created by pair 
production involving background $\gamma$-ray photons interacting with the high 
magnetic field. The pairs so created are accelerated in opposite directions 
to relativistic energies by the large potential difference across the gap. 
These highly energetic particles produced in the vacuum gap emit further 
$\gamma$-rays via curvature radiation and/or inverse compton scattering, 
giving rise to a cascade of $e^-e^+$ pairs which discharges the potential 
difference across the gap. The height of the gap region $h$ stabilizes at 
the mean free path length of the $\gamma$-ray photons and is given by RS75 as:\\
$h = 5\times10^3~b^{-4/7}~P^{1/7} \dot{P}_{-15}^{-2/7} {\rho_{6}}^{2/7}$ cm.\\
Here $\rho_{6}$ = $\rho$/10$^6$ cm, where $\rho$ is the radius of curvature of 
the multipolar field lines in the gap region. The potential difference across
the vacuum gap is $\Delta V = \Omega B h^2/c$ which, according to RS75, reduces
to:\\
$\Delta V = 5\times10^9~b^{-1/7} P^{-3/14} \dot{P}_{-15}^{-1/14} {\rho_{6}}^{4/7}$ \mbox{Volts}.\\
The discharge of the vacuum gap takes place through a series of sparks which 
deposit highly energetic singly charged primary particles beyond the gap 
region. According to the ``sparking'' model~\citep{gil00} the typical diameters
of the sparks are $\sim h$. The primary particles have maximum energies 
specified as $\gamma_b = e\Delta V /mc^2 \sim$ 3$\times $10$^6$ for typical 
pulsar parameters. Due to partial shielding of the acceleration region by 
thermionic ions, the gap potental drops to $\Delta V = \eta {\Delta V}_{max}$ 
\citep{gil03}, and the particle densities are given as $n_b = \eta n_{GJ}$, 
where 0 $ < \eta <$ 1 is the screening factor. Once outside the gap the primary
particles cease to accelerate but continue emitting curvature radiation with 
characteristic photon energies given by $\hbar \omega_c = \frac{3}{2} \gamma_
b^3 \hbar c/\rho$. The photons continue pair production resulting in a cloud 
of secondary $e^-e^+$ plasma with typical energies $\gamma_p = \hbar 
\omega_c/2mc^2 \sim$ 10-1000 and density $n_p = \chi n_{GJ}$, where $\chi 
\sim$ 10$^4$ is the Sturrock multiplicative factor \citep{stu71}. The spark 
discharge timescale in the vacuum gap is a few tens of $\mu$seconds, and this 
results in overlapping clouds of outflowing secondary plasma penetrated by 
the primary particles. Thus to summarize, the pulsar magnetosphere above the 
polar cap consists of a series of secondary plasma clouds, each corresponding 
to a ''spark" in the vacuum gap, penetrated by primary particles, moving 
outward along the magnetic field.

The principal predicament in explaining the radio emission from pulsars is 
its coherent nature evidenced by high brightness temperatures. This presages 
the presence of a bunching mechanism resulting in a large number of 
charged particles radiating simultaneously in phase. There are two possible 
regions where instabilities in the outflowing plasma can lead to charge 
bunching. In the first case particles of different momenta in the overlapping 
clouds undergo two stream instability giving rise to strong electrostatic 
Langmuir waves around heights of $\sim$ 50$R_S$. The Langmuir waves are 
modulationally unstable, and nonlinear plasma processes results in bunching 
of charged solitons~\citep{mel00}. The radio emission from the main pulse 
originates around these heights {\citep{ran93a,ran93b,kij98,kij03,mit02}} 
and is explained as curvature radiation from the charged solitons 
{\citep{glm04}}.In the second case the naturally developing electromagnetic
modes in the clouds of secondary plasma undergo negative absorption from the 
highly energetic primary paricles via the cyclotron resonance instability
\citep{kaz87,kaz91,lyu99} with the possibility of coherent radio emission. 
Such an emission near the light cylinder will likely be spread out over 
a large part of the pulsar's rotation cycle and is a likely source of 
off-pulse emission. In the next sections we examine the conditions necessary 
for the development of the cyclotron resonance instability near the light 
cylinder and its viability in explaining the detected off-pulse emission.

\section{\bf Conditions for emission near the light cylinder}
The magnetic field near the outer magnetosphere becomes strictly dipolar 
field due to the much faster decay of the higher multipoles. The magnetic 
field and plasma density, within the dipolar structure, decrease with distance 
from neutron-star surface as $(R_S/R)^3$. The high magnetic field near 
the neutron-star surface constrains the outflowing plasma to move along the 
field lines. This condition is relaxed near the light cylinder where the 
magnetic field is much weaker and the particles are able to gyrate. Within
the outflowing plasma various electromagnetic(EM) propagating modes can be 
generated. The extra-ordinary mode is one such example which is a transverse 
EM wave capable of propagating in the plasma and escaping the plasma region. 
The dispersion relation for the extra-ordinary waves is given as~\citep{kaz91}:
\begin{equation}
\omega = kc(1 - \delta),~~~\delta = {\displaystyle \frac{\omega_{p}^{2}}{4\omega_{B}^{2}\gamma_{p}^{3}}}
\label{eqn_1}
\end{equation}
Here $\omega_B$ is the cyclotron frequency and $\omega_p$ the plasma frequency.
\begin{equation}
\omega_{B} = {\displaystyle \frac{eB}{m_{e}c}};~~~~\omega_{p} = {\displaystyle \left(\frac{4\pi n_{p}e^{2}}{m_{e}}\right)^{1/2}}
\label{eqn_2}
\end{equation}
The amplitude of the extra-ordinary wave should grow as it propagates through
the plasma if it is to escape the plasma region. This amplification is provided 
by the high energy primary plasma through the cyclotron resonance instability. 
The primary particles near the light cylinder are capable of resonating with 
the extra-ordinary wave. The particles in addition also undergo curvature drift 
with a drift velocity $u_d$ = $\gamma c^2$/$\omega_B \rho$. The resonance 
condition for the instability is given as: 
\begin{equation}
\omega(\mathbf k)-k_{\|}v_{\|}-k_{\bot}u_d +{\displaystyle \frac{\omega_{B}}{\gamma}} = 0.
\label{eqn_3}
\end{equation}
The dispersion relation [eq.(\ref{eqn_1})] and the resonance condition 
[eq.(\ref{eqn_3})] gives the resonance frequency to be
\begin{equation}
\omega_0 = {\displaystyle \frac{\omega_{B}}{\gamma_{res}\delta}}.
\label{eqn_4}
\end{equation}
Here $\gamma_{res}$ = $\gamma_b$. In addition to the excitation of resonance 
frequency, another factor that is crucial for the emergence of the emission is 
the growth rate $\Gamma$ specified as~\citep{kaz91}:
\begin{eqnarray}
\Gamma = \pi{\displaystyle \frac{\omega_{p,res}^{2}}{\omega_0\gamma_{T}}}~~~~~~~u_d^2/2c^2\delta \ll 1 \nonumber \\
\Gamma = \pi{\displaystyle \frac{\omega_{p,res}^{2}}{2\omega_0\gamma_{T}}} {\displaystyle \frac{u_d^2}{c^2\delta}}~~~~~~~u_d^2/2c^2\delta \gg 1 
\label{eqn_5}
\end{eqnarray}
$\gamma_T$ is the thermal spread in the primary particle distribution. The 
necessary conditions for the electromagnetic emission due to the cyclotron 
resonance instability are:
\begin{enumerate}
\item The growth factor should be large $\Gamma\tau >$ 1, where $\tau \sim$ 
$R_{LC}/c$ = $P/2\pi$ is the growth time, i.e. the duration before the waves 
escape the light cylinder.
\item The resonance frequency should not exceed the damping frequency, 
$\omega_0 < \omega_1$ = $2\gamma_p \omega_B$.
\end{enumerate}

\subsection{Parametric representation}
\begin{figure*}
\includegraphics[angle=0,scale=1.3]{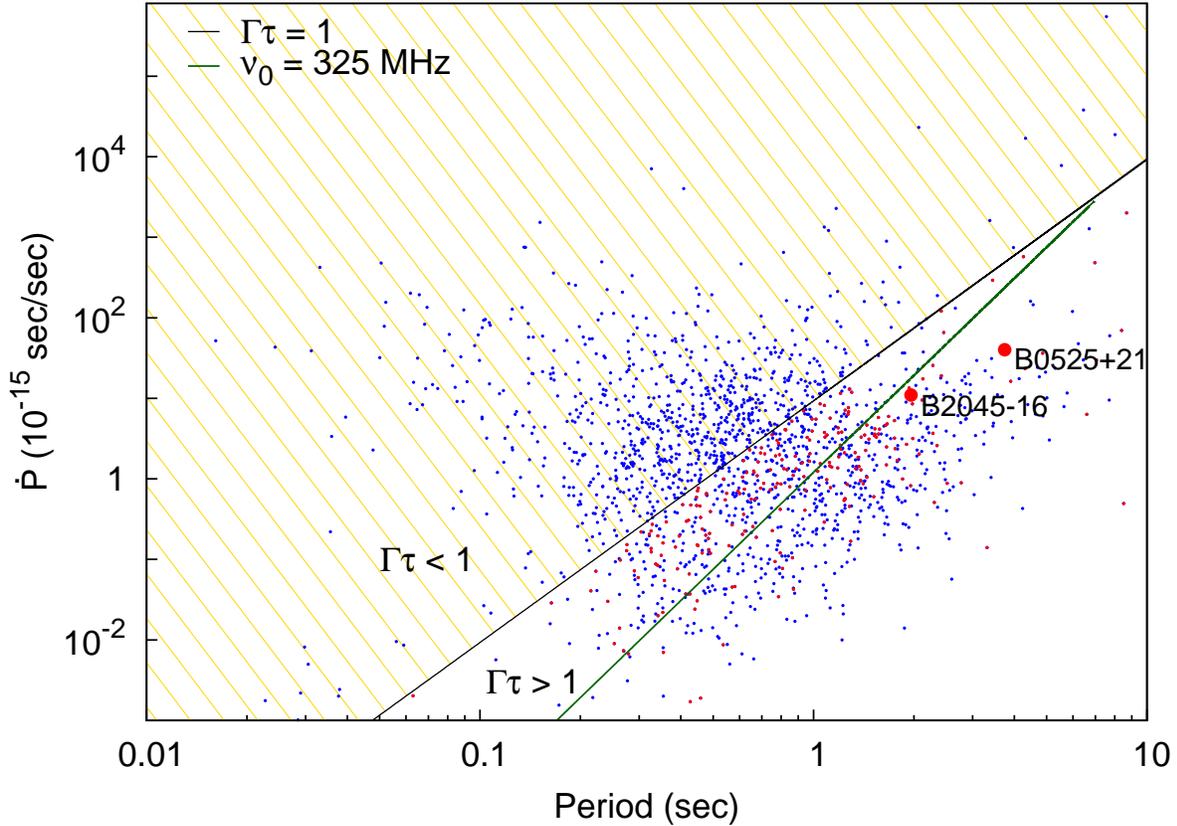}
\caption{The condition for cyclotron resonance instability for the pulsar 
population (red and blue points) is represented in the $P\dot{P}$ diagram with 
the relevant pulsar parameters given in \S 3.2. The solid black line 
corresponds to eq.(\ref{eqn_16}) where the growth factor $\Gamma\tau$ = 1. 
The instability condition develops in the pulsar population where 
$\Gamma\tau >$ 1 [eq.(\ref{eqn_16})]. The region where $\Gamma\tau <$ 1
(shaded region) represents the pulsars (blue points in shaded region) where 
radio emission in outer magnetosphere cannot originate. For the pulsar 
parameters assumed for our calculations (\S 3.1 and 3.2) the resonance 
frequency of emission at 325 MHz [eq.(\ref{eqn_14})] is shown as the green 
line. In addition to satisfying the instability condition (region where 
$\Gamma\tau >$ 1) the possibility of detecting radio emission depends on the 
flux density and the efficiency of energy in the plasma being converted into 
radio emission. Using an upper detection limit of 1 mJy for current radio 
telescopes at the sub GHz range, we estimate the minimum observable luminosity 
$L_{obs}$ for each pulsar using eq.(\ref{eqn_20}). Assuming the efficiency 
of the emission mechanism to be about 5\% of $L_p$ [given by eq.(\ref{eqn_19})]
we find a sub-sample (red points) where the emission mechanism presented here 
can give rise to detectable off-pulse emission (i.e. 0.05$L_p > L_{obs}$ and 
$\Gamma\tau >$ 1). The two pulsars PSR B0525+21 and  PSR B2045--16, where 
off-pulse emission is observed, lie in the favorable region for cyclotron 
resonance instability.
\label{fig}}
\end{figure*}

We assume the emission to manifest at the light cylinder, $R_{LC} = c/\Omega 
\approx 4.8 \times 10^{9} P$~cm, ensuring maximum growth of the resonant waves.
We determine a representation of the conditions explained above in terms of 
the basic parameters of the pulsar and the outflowing plasma. The magnetic 
field is given as:
\begin{eqnarray}
 B & \approx & 10^{12} (P\dot{P}_{-15})^{1/2}(R_S/R)^3~~G \nonumber \\
   & \approx &  9~(\dot{P}_{-15}/P^5)^{1/2}~~G
\label{eqn_6}
\end{eqnarray}
Here $R_S \sim$ 10$^6$ cm and $B$ is represented at the light cylinder. The 
Goldreich-Julian density on the neutron star surface is given as 
\begin{eqnarray}
 n_{GJ} & = & -({\bf{\Omega\centerdot B}})/2\pi ce \nonumber \\
        & \approx &  6.9 \times 10^{10}~({\dot{P}}_{-15}/P)^{1/2}~~\mbox{cm}^{-3}
\label{eqn_7}
\end{eqnarray}
Using eq.(\ref{eqn_6}) and (\ref{eqn_7}) and appropriate distance scaling,
the various plasma properties at the light cylinder are as follows.
\begin{eqnarray}
\omega_p^2 & \approx & 4.1\times10^9~\chi\left({\displaystyle\frac{\dot{P}_{-15}}{P^7}}\right)^{1/2}  \nonumber \\
\omega_{p,res}^2 & \approx & 4.1\times10^9~\eta\left({\displaystyle\frac{\dot{P}_{-15}}{P^7}}\right)^{1/2}  \nonumber \\
\omega_B & \approx & 3.3\times10^8\left({\displaystyle\frac{\dot{P}_{-15}}{P^5}}\right)^{1/2}  \nonumber \\ 
\delta & \approx & 9.6\times10^{-9}\left({\displaystyle\frac{\chi}{\gamma_p^3}}\right)\left({\displaystyle\frac{P^3}{\dot{P}_{-15}}}\right)^{1/2}
\label{eqn_8}
\end{eqnarray}
The frequency of the emitted waves and the damping frequency follows from 
eq.(\ref{eqn_8}).
\begin{eqnarray}
\omega_0&\approx&3.4\times10^{16}\left({\displaystyle\frac{\gamma_p^3}{\chi\gamma_{res}}}\right)\left({\displaystyle\frac{\dot{P}_{-15}}{P^4}}\right) \nonumber \\
\omega_1&\approx&6.5\times10^8~\gamma_p\left({\displaystyle\frac{\dot{P}_{-15}}{P^5}}\right)^{1/2}
\label{eqn_9}
\end{eqnarray}
In order to determine the growth condition we need to look into the relevance 
of the curvature drift. At the light cylinder $\rho \sim R_{LC}$, hence the 
drift velocity and drift condition are given as
\begin{eqnarray}
u_d \approx 5.7\times10^2~\gamma_{res}\left({\displaystyle\frac{P^3}{\dot{P}_{-15}}}\right)^{1/2} \nonumber \\
{\displaystyle\frac{u_d^2}{2c^2\delta}} \approx 1.9\times10^{-8}\left({\displaystyle\frac{\gamma_{res}^2\gamma_p^3}{\chi}}\right)\left({\displaystyle\frac{P^3}{\dot{P}_{-15}}}\right)^{1/2}
\label{eqn_10}
\end{eqnarray}
For typical pulsar parameters it can be shown that $u_d^2/c^2\delta \gg$ 1,
establishing the importance of curvature drift. The growth factor using the 
second part of eq.(\ref{eqn_5}) is   
\begin{equation}
 \Gamma\tau \approx 1.2\times10^{-15}\left({\displaystyle\frac{\eta\gamma_{res}^3}{\gamma_T}}\right)\left({\displaystyle\frac{P^3}{\dot{P}_{-15}}}\right)
\label{eqn_11}
\end{equation}
Finally the limits on frequency, $\omega_0 < \omega_1$, and growth factor, 
$\Gamma\tau >$ 1, constrains the pulsar and plasma parameters.
\begin{eqnarray}
\left({\displaystyle\frac{\chi\gamma_{res}}{\gamma_p^2}}\right)\left({\displaystyle\frac{P^3}{\dot{P}_{-15}}}\right)^{1/2} > 5.2\times10^7 \nonumber \\
\left({\displaystyle\frac{\eta\gamma_{res}^3}{\gamma_T}}\right)\left({\displaystyle\frac{P^3}{\dot{P}_{-15}}}\right) > 8.3\times10^{14}
\label{eqn_12}
\end{eqnarray}
The conditions for the emission are shown in Fig.~\ref{fig} for the pulsar
population in the $P\dot{P}$ diagram.

\subsection{Application to off-pulse emission}
We now look into the application of the parametric formulation for a specific 
set of pulsar parameters and apply them to two pulsars with detected off-pulse 
emission. The pulsar parameters used for our studies are, $\gamma_p$ = 10;
$\chi$ = 10$^4$; $\gamma_{res}$ = 2$\times$10$^6$; $\gamma_T$ = 10$^2$; $\eta$ 
= 10$^{-1}$. We look at the various conditions as explained above.
\begin{equation}
{\displaystyle\frac{u_d^2}{2c^2\delta}} \approx 7.6\times10^2\left({\displaystyle\frac{P^3}{\dot{P}_{-15}}}\right)^{1/2}
\label{eqn_13}
\end{equation}
\begin{equation}
\nu_0 = \omega_0/2\pi \approx 270\left({\displaystyle\frac{\dot{P}_{-15}}{P^4}}\right)~\mbox{MHz}
\label{eqn_14}
\end{equation}
\begin{equation}
\nu_1 = \omega_1/2\pi \approx 1.03\left({\displaystyle\frac{\dot{P}_{-15}}{P^5}}\right)^{1/}~\mbox{GHz}
\label{eqn_15}
\end{equation}
\begin{equation}
\Gamma\tau \approx 9.3\left({\displaystyle\frac{P^3}{\dot{P}_{-15}}}\right)
\label{eqn_16}
\end{equation}

{\bf B0525+21:} $P$ = 3.7455 sec; $\dot{P}_{-15}$ = 40.05; \\
The limits expressed in eq.(\ref{eqn_12}) reduce to\\
$\left({\displaystyle\frac{\chi\gamma_{res}}{\gamma_p^3}}\right)\left({\displaystyle\frac{P^3}{\dot{P}_{-15}}}\right)^{1/2} = 2.3\times10^8>5.2\times10^7$
$\left({\displaystyle\frac{\eta\gamma_{res}^3}{\gamma_T}}\right)\left({\displaystyle\frac{P^3}{\dot{P}_{-15}}}\right) = 1.05\times10^{16}>8.3\times10^{14}$\\

{\bf B2045--16:} $P$ = 1.9616 sec; $\dot{P}_{-15}$ = 10.96; \\
The limits expressed in eq.(\ref{eqn_12}) reduce to\\
$\left({\displaystyle\frac{\chi\gamma_{res}}{\gamma_p^3}}\right)\left({\displaystyle\frac{P^3}{\dot{P}_{-15}}}\right)^{1/2} = 1.7\times10^8>5.2\times10^7$
$\left({\displaystyle\frac{\eta\gamma_{res}^3}{\gamma_T}}\right)\left({\displaystyle\frac{P^3}{\dot{P}_{-15}}}\right) = 5.5\times10^{15}>8.3\times10^{14}$\\

The pulsar parameters for the two pulsars have been obtained from the ATNF 
pulsar database~\citep{man05}.

As shown in Fig.~\ref{fig}, the two pulsars B0525+21 and B2045--16 which 
show the presence of off-pulse emission~\citep{bas11,bas12} lie in the regime 
where cyclotron resonance instability is likely to emit observable coherent 
radio emission. The resonance frequency of emission at 325 MHz given by 
eq.(\ref{eqn_14}) is shown as a green line in Fig.~\ref{fig}. 

In the analysis presented here we have made two basic assumptions that the
resonance condition develops at the light cylinder and the secondary plasma
is characterized by $\gamma_p$ = 10. These assumptions, although they 
demonstrate the viability of the emission mechanism and constrain the 
parameter space, do not put strict limits on the secondary plasma energy 
or location of emission in the pulsar magnetosphere. In one scenario for 
the same pulsar it is possible for the resonance condition to originate at 
a different height for a slightly different secondary plasma energy, i.e, 
lower energy particles will emit the same resonance frequency at a lower 
height. In an alternate picture the same energy particles will satisfy 
the resonance condition at a different height for a different resonance 
frequency, i.e, the resonance frequency will be higher at a lower height. 
In a typical pulsar all these scenarios are likely to exist, which would 
account for not only the wide range of frequencies likely to be emitted 
but also the spread of the detected signal over a wide range in the pulsar 
profile. However the emission conditions can only develop in the outer 
magnetosphere near the light cylinder, which in conjunction with the damping
frequency puts a limit on the maximum frequency that can be emitted to be
around the GHz range.

\section{\bf Energetics}
In this section we look into the total energy available in the plasma and its
sufficiency in accounting for the detected off-pulse emission. The total 
available energy in the plasma when emitted as radio emission gives
\begin{equation}
L_{p} = \gamma_{res}mc^{3}n_{res}\beta_{p}~~~\mbox{erg}~s^{-1} 
\label{eqn_17}
\end{equation}
Here $\beta_{p}$ represents the cross-section of the open field lines,
$\beta_{p} = 6.6\times10^{8}P^{-1}(R/R_{s})^{3}$ cm$^{2}$. The luminosity
of the off-pulse emission is given as:
\begin{equation}
 L_{obs} = 4\pi D_{L}^{2}S_{\nu}\Delta\nu~\zeta^{-1}~~~\mbox{erg}~s^{-1}
\label{eqn_18}
\end{equation}
Here $D_{L}$ is the distance to pulsar; $S_{\nu}$ the observed off-pulse flux;
$\Delta\nu$ is the frequency range and $\zeta$ the fractional opening angle. 
Using $\gamma_{res}$=2$\times$10$^{6}$; $\eta$=10$^{-1}$; $S_{\nu}$=1 mJy; 
$\Delta\nu$=100 MHz; $\zeta$=10$^{-1}$ eq.(\ref{eqn_17}) and (\ref{eqn_18}) 
are expressed as
\begin{equation}
L_{p} \approx 2.2\times10^{29}\left({\displaystyle\frac{\dot{P}_{-15}}{P^3}}\right)^{1/2}
\label{eqn_19}
\end{equation}
\begin{equation}
L_{obs} \approx 1.2\times10^{27}D_{kpc}^2
\label{eqn_20}
\end{equation}
Applying the luminosity values to the two pulsars as in the previous section
we look into the relevance of the mechanism.\\
{\bf B0525+21:} $D_{kpc}$ = 2.28. $ L_{p}$ = 2.5$\times$10$^{29}$; $L_{obs}$ 
= 6.2$\times$10$^{27}$.\\
{\bf B2045--16:} $D_{kpc}$ = 0.95. $ L_{p}$ = 1.8$\times$10$^{29}$; $L_{obs}$ 
= 1.1$\times$10$^{27}$.\\
$L_{p} \gg L_{obs}$ for off-pulse emission in both pulsars. It is clear that 
the cyclotron resonance instability is an energetically viable mechanism for 
off-pulse emission in the two examples shown. 

\section{\bf Discussion}

In this work we have demonstrated the generation of radio emission due to 
the cyclotron resonance instability within the outer magnetosphere of pulsars 
making it a relevant candidate for off-pulse emission. The conditions for 
radio emission are dependent on the parameters of the plasma which are still 
poorly understood. The details of the calculations shown here will vary with
changes in plasma parameters although the basic physical processes would
still operate in pulsars. We can draw two primary conclusions about the nature 
and extent of the emission from these studies. Firstly, these conditions can 
only develop in a certain population of pulsars which have longer periods 
and/or smaller period derivatives. Secondly, there is an upper limit to the 
frequency that can be emitted constrained by the damping frequency. In 
addition we have also found an observational limit based on the detection 
capabilities of present day telescopes and energetics which show the emission 
from long period pulsars with very small period derivatives generally too weak 
to be observed.  The pulsars which follow the above criteria are shown as 
red points in Fig.~\ref{fig} (see figure caption for details). 

Further observational results involving a large sample of pulsars in addition 
to determining the temporal structure, polarization features and the spectral 
nature of the off-pulse emission will not only help in validating the 
mechanism, but will also help in refining its predictions.

\acknowledgments We thank the anonymous referee for constructive criticism
that helped to improve the paper. We would like to thank Joanna Rankin for 
carefully reading the manuscript and her helpful comments which improved the 
manuscript. We also thank Janusz Gil for his useful comments. This work was 
partially supported through the grant DEC-2012/05/B/ST9/03924 by the Polish 
National Science Centre.

\clearpage

\end{document}